\documentclass[aps,prl,floats,twocolumn,floatfix,amsmath,showpacs,unsortedaddress,superscriptaddress]{revtex4}

\usepackage[latin1]{inputenc}
\usepackage{SIunits}
\usepackage{graphicx}

\begin{document}

\title{Coherent Transport of Atomic Quantum States in a Scalable Shift Register}
\author{A. Lengwenus}
\author{J. Kruse}
\author{M. Schlosser}
\author{S. Tichelmann}
\author{G. Birkl}
\email{gerhard.birkl@physik.tu-darmstadt.de}%
\affiliation{Institut f\"ur Angewandte Physik,
        Technische Universit\"at Darmstadt, 64289 Darmstadt, Germany}

\date{\today}

\pacs{37.10.Jk, 42.50.Ct, 03.67.-a}

\begin{abstract}
We demonstrate the coherent
transport of two-dimensional (2D) arrays of small ensembles of neutral atoms 
in a shift register architecture based on 2D arrays of microlenses. 
We show the scalability of the transport process 
by presenting the repeated hand-over of atoms from
site to site.
We prove the conservation of coherence during transport, reloading, and a 
full shift register cycle.
This shows that the fundamental shift sequence can be cascaded and thus scaled to complex
and versatile 2D architectures 
for atom-based quantum information processing, 
quantum simulation, and the investigation of quantum degenerate gases. 
\end{abstract}

\maketitle 

\begin{figure}[b]
\includegraphics[width=1.0\linewidth]{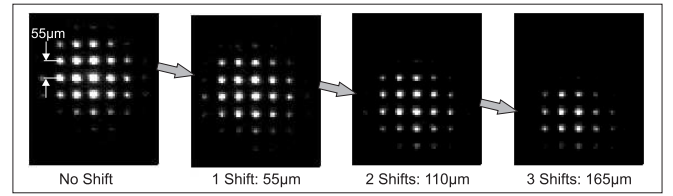}
\caption{Fluorescence images of 2D arrays 
of trapped atoms in a shift register based on 2D arrays of microlenses.
Images are taken after 0, 1, 2, and 3 consecutive shift sequences of 55 $\mu m$
transport distance each. 
\label{fig:array_sr}}
\end{figure}

\begin{figure}[b]
\includegraphics[width=0.7\linewidth]{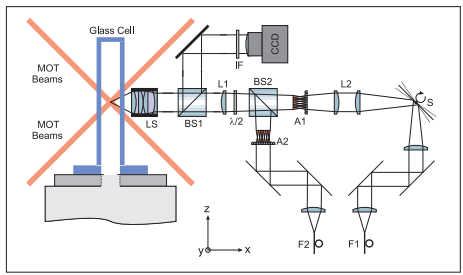}
\caption{(color online) Experimental setup.
Microlens array A1 is illuminated with light
delivered by optical fiber F1. The incident angle on A1 is
controlled by scanning mirror S in combination with the transfer
lenses L2. A second fixed array A2 is illuminated with light 
delivered by fiber F2. The focal planes of both arrays
are superimposed by beamsplitter BS2 and transferred by lens
L1 and lens system LS into the MOT region. Detection of
fluorescence light is performed through LS and beamsplitter BS1 by 
a CCD camera with interference filter IF to block straylight.
\label{fig:setup}}
\end{figure}

\begin{figure}[t]
\includegraphics[width=0.8\linewidth]{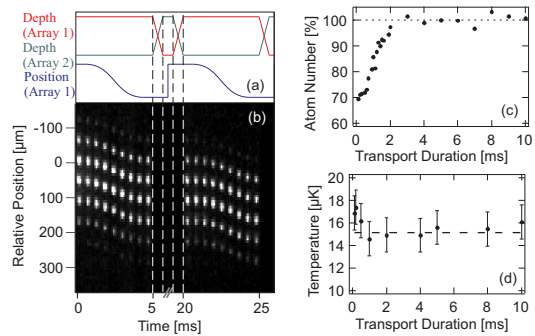}
\caption{(color online) (a) Timing sequence for depth and position
of the two trap arrays constituting a shift register. (b) Fluorescence images
of atoms in the central column of the register of Fig. 1 during two consecutive
shift cycles. No images are shown for the phases of loading and reloading between the arrays 
and the phase of returning array A1 to its start position (\unit{5}{\milli\second} each). (c) Atom number and (d) temperature in the central trap at the end of a
single transport sequence of 55 $\mu m$ distance
as a function of the transport duration. 
The dashed line in (d) shows the temperature for a fixed trap. 
Each data point is averaged 15 times. 
\label{fig:ribbon}}
\end{figure}

In many of the recent advances in the investigation
of quantum degenerate gases and neutral atom quantum information processing, 
versatile architectures for the coherent storage and transport 
of atomic quantum systems based on optical dipole potentials play a crucial role
\cite{2001Science..293i278,2003Nature..425,PhysRevLett.93.150501,2005Bloch_optical_lattices,
2006Nature..442,2007Nature..448,2001PhysRevLett.88.020401,2002PhRvL..89i7903D,
2006PRL..96i063001,2007natphy..3i697,2007PhRvA..75d0301J}.
In specific, the application of standing-wave configurations (optical lattices)
\cite{2001Science..293i278,2003Nature..425,PhysRevLett.93.150501,
2006Nature..442,2007Nature..448,2005Bloch_optical_lattices}
and single or multiple focused laser beams 
\cite{2001PhysRevLett.88.020401,2002PhRvL..89i7903D,2006PRL..96i063001,2007natphy..3i697,2007PhRvA..75d0301J}
has led to significant progress in the manipulation of atomic qubit states
for quantum information processing. 
In our work, we focus on the implementation of geometries
based on microfabricated optical elements \cite{2002PhRvL..89i7903D,2001OptComm,multirefapq,
PhysRevA.81.060308}. 
This approach allows us to develop flexible and integrable configurations for quantum state
storage and manipulation, simultaneously
targeting the important issues of single-site addressing and scalability, essential to most  
architectures for quantum information processing \cite{2000ForPh..48..771D} and 
quantum simulation.\\ 
The scalable shift register presented here is an all optical
device which offers precise control of the transport of trapped
neutral atoms in a two-dimensional (2D) architecture. 
The atoms are localized in miniaturized arrays of
dipole potentials created by 2D microfabricated 
lens structures \cite{2001OptComm,2002PhRvL..89i7903D}. 
The shift operation is based on consecutive
loading, moving, and reloading of two independently controllable arrays of
traps. Figure \ref{fig:array_sr} shows a 2D register of about 25 atom samples
(each able to carry one quantum bit (qubit) \cite{PhysRevA.81.060308})
detected by collecting the fluorescence light emitted by
the atoms when illuminated with a resonant laser pulse after 0, 1, 2, and 3 consecutive shift
sequences.
The shift register
allows for atom transport over macroscopic distances and at the same time
for controlled atom-atom approach with sub-micrometer precision which is necessary for 
the implementation of two-qubit quantum gates
\cite{1999PhRvL..82.1060B,1999PhRvL..82.1975J,2000PhRvL..85.2208J,2000PhRvA..61b2304C,
2002PhRvA..66d2317E,2003PhRvL..90n7901M,2007PhRvL..98,
PhysRevLett.104.010502,PhysRevLett.104.010503}. 
Moreover, it can serve as a two-dimensional
quantum memory to archive and retrieve quantum information, or sequentially 
shuffle quantum information through complex architectures.
Conservation of coherence of the quantum states and
adiabaticity during transport are essential requirements that will be 
addressed in this article. 
An extension of the results for one-dimensional (1D) shift operations presented here 
to 2D is straightforward.
\\
%
%
The experiments presented here are performed with $^{85}$Rb atoms
inside a glass cell based vacuum system (Fig.
\ref{fig:setup}). Each experimental sequence is initiated by preparing an
ensemble of about $10^{6}$ atoms in a standard retroreflected
magneto-optical trap (MOT). The atoms are further cooled by 
optical molasses to approximately \unit{10}{\micro\kelvin} before
being partially transferred into a superimposed 2D register of dipole traps. 
The traps are
created by illuminating a subset of a two-dimensional array (A1) of $50 \times 50$
microfabricated refractive lenses with light far red-detuned 
from the
D1 and D2 transitions of Rb.
The microlenses have a diameter of \unit{100}{\micro\metre}, a pitch
of \unit{125}{\micro\metre}, and a focal length of
\unit{1}{\milli\metre}. The focal plane of the array is relayed
into the glass cell using a telescope which consists of an
achromatic lens (L1, $f=\unit{80}{\milli\metre}$) and a diffraction
limited lens system (LS, $f=\unit{35.5}{\milli\metre}$, $NA=0.29$). 
The demagnification results in traps with a
separation of $a=\unit{55}{\micro\metre}$ and a measured waist of
\unit{3.8}{\micro\metre} ($1/e^{2}$ radius). 
Illuminating the microlens array with a laser beam at
\unit{805}{\nano\metre} wavelength, a
power of \unit{275}{\milli\watt}, and a $1/e^2$ radius of \unit{450}{\micro\metre}, 
results in a two-dimensional array of traps with a 
power of \unit{5.7}{\milli\watt} 
and a depth of $k_{\textrm{B}} \times \unit{430}{\micro\kelvin}$
for the central trap. 
Here, the vibrational frequencies are $\Omega_{\textrm{r}}=2\pi\times\unit{17}{\kilo\hertz}$
for the radial and
$\Omega_{\textrm{a}}=2\pi\times\unit{820}{\hertz}$ for the axial
direction. 
About 200 atoms with a temperature of $\unit{15\pm1.5}{\micro\kelvin}$ (measured
by a time-of-flight technique) are trapped in the central trap.
Because of the Gaussian
profile of the laser beam illuminating the microlens array, the outlying
traps are shallower.
The number of traps loaded (here about $5 \times 5$) depends on this beam size but 
also on the size of the MOT. The 
lifetime of the atoms in the traps is on the order
of \unit{0.5}{\second} which is mainly limited by collisions
with background gas atoms.
Atom detection is achieved by resonance fluorescence imaging of the atom 
distribution using the MOT beams for illumination and collecting the fluorescence 
light with an EMCCD camera through lens system LS and beamsplitter BS1.\\
%
%
To move the traps, we vary the incident angle on microlens
array A1 by a feedback-controlled scanning mirror S which
deflects the incoming beam. 
The pivot point of the beam on the scanning
mirror is imaged onto the microlens array by 
telescope L2 with unity magnification.
This causes the foci of the array to shift laterally within 
the focal plane as a function of the angle of
the scanning mirror.
It is straightforward to shift the
array by a distance of the full trap separation of
\unit{55}{\micro\meter}. Moving significantly more than this
distance results in strong deformations of the trapping potentials
by lens abberations due to the skewed
illumination of the microlenses.
The angular reproducibility of the scanner is
better than \unit{22}{\micro\rad} which implies that the trap position can be
controlled to better than
\unit{10}{\nano\meter}. \\
A complete shift register sequence consists of consecutive loading, moving, and
reloading of two independently controllable arrays of dipole traps. 
The fixed focal structure of microlens
array A2 (identical to lens array A1) is combined with the movable focal structure of 
array A1 by beamsplitter BS2. 
For shift and reloading durations of \unit{5}{\milli\second}, the timing sequence
for the potential depths of arrays A1 and A2  and for the position of array A1 are shown in Fig.
\ref{fig:ribbon} (a). The fluorescence images in Fig.
\ref{fig:ribbon} (b) show the central column
of the array of Fig. \ref{fig:array_sr} as a function of time
during two consecutive shift cycles. 
The shift operation is comparable to a bucket chain: while
A2 is switched off initially, we load atoms from the MOT into A1 at $-a/2$ and shift it by one full trap separation from $-a/2$ to $+a/2$ in
a few {\milli\second}. We transfer the atoms to A2 (illuminated under normal angle but
laterally displaced by $+a/2$) by
rising the intensity in A2 while ramping it down in A1.
Then the scanning mirror is returned to its
initial position which superimposes the next row of traps of A1 with the traps of A2. 
To complete a shift cycle, the atoms are reloaded from A2 to A1,
and the next fully identical shift cycle can begin. 
The number of achievable shift sequences 
is only limited by the size of the illuminated trap array. 
The $5 \times 5$ trap array in this realization 
for example allows for 5 sequences and gives a final transfer distance of $\unit{275}{\micro\metre}$. 
For sufficient laser power, fully exploiting the here available set of $50 \times 50$ microlenses 
allows for atom transport over many trap separations.
\\
%
%
Minimizing atom loss and heating are essential for the shift register but
e.g. also for specific realizations
of quantum gates where atoms have to be brought close together 
and separated again \cite{1999PhRvL..82.1060B,1999PhRvL..82.1975J,2000PhRvL..85.2208J,2000PhRvA..61b2304C,
2002PhRvA..66d2317E,2003PhRvL..90n7901M,2007PhRvL..98,
PhysRevLett.104.010503,PhysRevLett.104.010502}.
For this reason, we investigated in detail heating and atom loss during
individual transport operations.
Figure \ref{fig:ribbon} shows the atom
number (c) and temperature (d) at the end of single transport sequences 
with varying duration. The transport
distance is one full trap separation. 
Typical initial temperatures are $\unit{15\pm1.5}{\micro\kelvin}$.
For transport durations above
\unit{1}{\milli\second} we measure no increase in temperature and for 
transport durations above \unit{2}{\milli\second}
no atom loss. Heating and atom loss for faster transport
arise from technical
limitations of the scanning mirror: for short scan
times, the mirror has to be strongly accelerated and decelerated,
causing the mirror
to overshoot its final position because of its inertia.
%
%
This
results in oscillations around the final position 
which cause
resonant excitation of the vibrational levels and as a consequence
heating and atom loss. Optimization allowed us to minimize this effect 
by smoothing 
the ramps for acceleration and deceleration as shown 
in the trace for the position of array A1 in Fig. \ref{fig:ribbon} (a) and thus to
reduce the required scan duration to \unit{2}{\milli\second}.
\\
We also investigated atom loss and heating for a shift register
consisting of several shift cycles.
After a sequence of four cycles 
(Fig. \ref{fig:array_sr}), 
heating was measured to be below \unit{2}{\micro\kelvin}, 
which is comparable to our measurement uncertainty. 
On the other hand, atom loss on the order of 
20 \% was encountered for each loading from A1 to A2, whereas no loss was observed for
reloading from A2 to A1. In a separate series of measurements 
we found that this was caused by the fact that 
loading and reloading have not been fully symmetrical operations since A1
was displaced to $+a/2$ by tilted illumination whereas A2 was laterally moved 
to $+a/2$ but illuminated normal.
This leads to slightly tighter traps in A2 as compared to A1. 
Minimizing this mismatch could be achieved through symmetrizing the loading and reloading
processes by having both arrays displaced by $|a/2|$ with opposite sign through tilted illumination 
during the loading and reloading phases.
This
allows us to reload atoms in both directions without detectable loss.\\
%
%
For quantum information processing in this architecture,
also the coherence of superpositions of 
quantum states has to be preserved during transport, 
reloading, and the full shift cycle. In our approach, qubit states are
represented by hyperfine substates of the $5\textrm{S}_{1/2}$
ground state of $^{85}\textrm{Rb}$.
To be
insensitive to fluctuations of magnetic fields to first order, we prepare the
atoms in the clock state ($F=2$, $m_F=0$) which we coherently couple to
the second clock state ($F=3$, $m_F=0$) 
with a phase-locked diode laser system. 
The typical duration of an
applied $\pi$-pulse is \unit{210}{\micro\second}. 
State-selective detection is performed by removal of the atoms in $F=3$
and subsequent detection of the remaining atoms in $F=2$.\\ 
%
\begin{figure}[tb]
\includegraphics[width=0.8\linewidth]{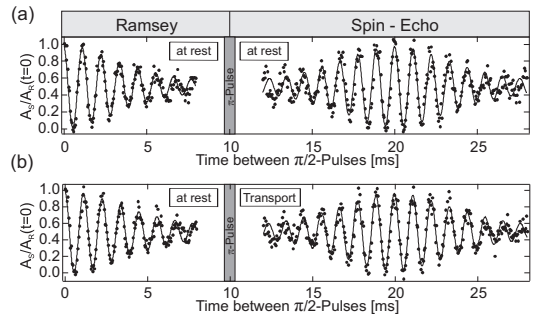}
\caption{Spin-echo measurements for atoms at rest (a, right) and 
atoms transported over a distance of 55 $\mu m$ (b, right) in the 
central trap of a two-dimensional trap array. 
In both cases, the spin-echo signal ($A_S$) is normalized to the initial amplitude
($A_R$(t=0)) of a Ramsey signal for atoms at rest (left). 
There is almost negligible additional loss of signal contrast
through atom transport. Each data point is averaged 5 times.
\label{fig:sr_echo}}
\end{figure}
%
We analyze the influence of atom transport
on dephasing and decoherence by applying Ramsey and spin-echo
methods.
The trapping laser light is tuned to a wavelength of \unit{815}{\nano\meter} with a
power of \unit{150}{\milli\watt} distributed over a beam with 
a $1/e^2$ radius of \unit{520}{\micro\metre}. For the central trap,
this gives a power of \unit{2.3}{\milli\watt} 
and a trap depth of
$k_B\times\unit{110}{\micro\kelvin}$. 
As the
atom ensemble in each trap is thermal and thus distributed over
a range of vibrational 
levels, each atom incurs a slightly different ac-Stark shift.
This leads to inhomogeneous dephasing of the Ramsey signal already 
for atoms at rest (Fig. \ref{fig:sr_echo} (left)) with 
a time constant of about \unit{5}{\milli\second}.
To compensate for this,  
we implemented a spin-echo technique
which reverses inhomogeneous dephasing. This allows 
to directly measure the combined time constant $T_{2}'$ of
homogeneous dephasing and decoherence 
and thus to compare the behavior of atoms at rest and atoms transported.
For atoms at rest (Fig. \ref{fig:sr_echo} (a, right)), 
a $\pi/2$-pulse at $t=0$ is followed by a first period
of free evolution. 
Rephasing is induced by
a $\pi$-pulse after $t_{\pi}$. After an additional period of 
free evoluton with variable duration, the
sequence is completed by a second $\pi/2$-pulse and
detection of the atoms in $F=2$. The maximum
amplitude of the echo signal occurs at $t=2t_{\pi}$
($2t_{\pi}=\unit{20}{\milli\second}$ in Fig. \ref{fig:sr_echo}).
For the case of transported atoms (Fig. \ref{fig:sr_echo} (b, right)), 
transport over a distance of $\unit{55}{\micro\metre}$
takes place during the first \unit{2}{\milli\second} of the first
phase of free evolution with no transport during the second phase. 
Additional dephasing and decoherence caused by atom transport during the first
phase are not
compensated during the second phase and should cause a 
reduction of the signal amplitude at $t=2t_{\pi}$. Fig. \ref{fig:sr_echo} shows that this
effect is almost neglible in our system.\\
%
\begin{figure}[b]
\includegraphics[width=0.80\linewidth]{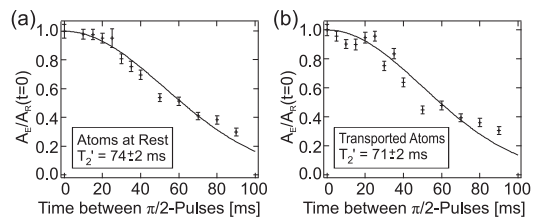}
\caption{Determination of the time constant $T'_2$ for homogeneous dephasing
and decoherence for atoms at rest (a) and atoms transported over 55 $\mu m$ in 2 ms (b). 
Atom transport has an almost negligible effect on dephasing and decoherence.
\label{fig:sr_coherence}}
\end{figure}
%
%
\begin{figure}[b]
\includegraphics[width=0.6\linewidth]{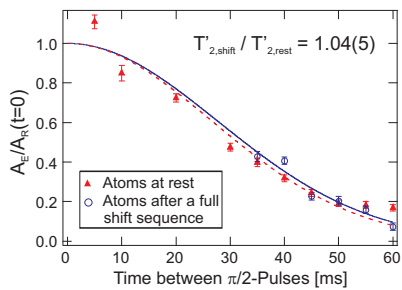}
\caption{(color online) Determination of the time constant $T'_2$ for homogeneous dephasing
and decoherence for a full shift register cycle. 
The signal contrast for atoms at rest (triangles) and atoms after the shift sequence (circles)
are shown.
The shift sequence has no measurable effect on dephasing and decoherence.
Due to imperfect experimental conditions, both absolute time constants $T'_2$ are smaller than
in Fig. \ref{fig:sr_coherence}.
\label{fig:Total_Transfer}}
\end{figure}
%
For a quantitative investigation, 
Fig. \ref{fig:sr_coherence} presents the signal contrast, i.e. the
maximum amplitude $A_{\textrm{E}}$ of the echo signal at $2t_{\pi}$ normalized to the
amplitude of the Ramsey signal $A_{\textrm{R}}(t=0)$ for a single central trap as a function
of $2t_{\pi}$ for
atoms at rest (a) and atoms transported over
$\unit{55}{\micro\metre}$ within \unit{2}{\milli\second} (b).
The loss in signal contrast is clearly non-exponential in both cases.
From a detailed analysis of external influences, 
we determine homogeneous dephasing due to irreversible
variations of the atomic resonance frequency 
to be the dominant cause for loss of contrast.
We identify heating due
to photon scattering from the trapping laser with a heating rate
too small to be directly observable 
in Fig. \ref{fig:ribbon} (d) to be the most likely cause for this. 
Following the calculations for 
homogeneous dephasing given in \cite{2005PhRvA..72b3406K}, the
signal contrast should be described by the Gaussian function
\mbox{$C(2t_{\pi})=C(0)\exp(-(2t_{\pi})^{2}/T_{2}'^2)$}
with time constant $T_{2}'$ for reduction of the initial contrast
to its $1/e$-value. 
The measurements in Fig. \ref{fig:sr_coherence} can be well fitted to \mbox{$C(2t_{\pi})$} (solid lines) which gives the time constants
\mbox{$T_{2, \textrm{rest}}'=\unit{74\pm2}{\milli\second}$} for atoms at rest,
\mbox{$T_{2, \textrm{trans}}'=\unit{71\pm2}{\milli\second}$} for atoms transported, and a
ratio \mbox{$T_{2, \textrm{trans}}'/T_{2, \textrm{rest}}'=0.96(4)$}, which is
consistent with 1 within our measurement uncertainty.
Thus, atom transport causes almost negligible additional dephasing and decoherence.\\ 
We have performed analogous measurements for the sequence of loading and reloading atoms from A1 to A2
and back to A1, and again found no significant decrease in $T_{2}'$.
Finally, we have investigated dephasing and decoherence for a full shift register cycle.
The cycle consists of preparing atoms in A1, transporting them over distance $a/2$, 
loading them from A1 to A2, and reloading them from A2 to A1. 
In Fig. \ref{fig:Total_Transfer}, the signal contrast for atoms at rest (triangles)
and atoms after the shift register cycle (circles) in a single central trap 
are presented as a function of $2t_{\pi}$. Fitting both data sets to \mbox{$C(2t_{\pi})$} allows 
to extract
the ratio
\mbox{$T_{2, \textrm{shift}}'/T_{2, \textrm{rest}}'=1.04(5)$}. 
Thus, also for the full shift register cycle, no additional dephasing or decoherence 
of internal-state superposition states occur within the measurement uncertainty.
This, by repetition, allows to build a complete shift register.\\
%
%
In summary, we have presented a novel shift register for atomic quantum systems
based on arrays of microfabricated lenses.
We have demonstrated that transport, reloading, and a full shift register cycle can
be performed with negligible atom loss, heating, or additional dephasing or decoherence.
This proves that the fundamental shift sequence can be cascaded and thus scaled to complex
and versatile 2D architectures allowing coherent quantum state storage and transport 
along complex and reconfigurable paths in 1D and - by simply upgrading
our 1D scanner to a 2D-version - also in 2D.
Together with parallelized site-selective single atom detection 
\cite{Single_Atoms_2010} and site-selective quantum-state manipulation \cite{PhysRevA.81.060308,PhysRevLett.104.010502,PhysRevLett.104.010503} novel
geometries for quantum informaton processing, quantum simulation, and multi-particle entanglement
become accessible.\\ 
We acknowledge financial support
by the DFG, 
by the European Commission (IP SCALA), 
by NIST (award 60NANB5D120),
and by the DAAD (contract 0804149).
\bibliography{sr08_30}
\bibliographystyle{apsrev}
\end{document}